# Report of the Notre Dame Contribution to the African School of Fundamental Physics and Applications 2022 in Gqeberha, South Africa

**Kenneth Cecire, Department of Physics and Astronomy**
**QuarkNet National Staff Teacher**
kcecire@nd.edu

**Shane Wood, Moundsview Public Schools and Notre Dame**
**QuarkNet National Staff Teacher**
swood5@nd.edu

**Submitted March 20, 2023**

**Abstract**
From November 26 to December 12, 2022, Shane Wood and Kenneth Cecire, QuarkNet staff members under the University of Notre Dame, traveled to South Africa as Lecturers in the African School of Fundamental Physics and Applications [1] (ASP) 2022, held at Nelson Mandela University in Gqeberha (Port Elizabeth) within the same calendar period. ASP is held every other year in a different African country for two or three weeks for African graduate and advanced undergraduate physics students to expose them to cutting-edge physics content and analysis techniques that may not be as available in their home institutions. Since 2016, there has been an outreach component consisting of the High School Teachers and Learners Programs. Cecire and Wood were facilitators of these programs and also acted as lecturers for two regular ASP classes. (We will use the term "students" to refer to these university students and "learners" to refer to high school students, following the practice of ASP.) They played a very active role and were quite busy during their two-week involvement. The mission was successful in terms of reaching teachers, learners, and students with new and exciting ideas and in terms of building strong collaborative relationships.

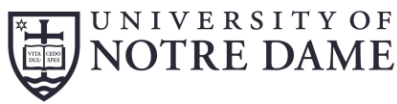
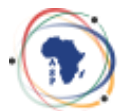

# Report on Notre Dame Contribution to the African School of Physics 2022

**Introduction**
QuarkNet [2] has been one of the main drivers of ASP outreach since the addition of this feature was requested by the government of Rwanda for ASP 2016 in Kigali [3]. Cecire has worked on this aspect of ASP since then and has seen it evolve and grow at the behest of ASP Director Ketevi Assamagan (who is now also a QuarkNet mentor at Brookhaven National Laboratory) and the International Organizing Committee.

The Teachers Program was held in the first week of ASP 2022, from November 28 to December 2, for 76 South African high school physics teachers, almost all from Eastern Cape Province of South Africa. The offerings for the teachers were divided between basic physics related to South African Department of Education Physical Science requirements [4] and contemporary physics activities with a stress on particle physics. Wood and Cecire facilitated these latter offerings, which were based on the QuarkNet Data Activities Portfolio.

In the second week, December 5 to 9, Wood and Cecire facilitated the Learners Program, with a three-hour workshop each morning for a different group of learners. These workshops focused, again, on contemporary physics based on the QuarkNet Data Activities Portfolio. Crucially, they also featured Q&A with ASP Lecturers, all physicists and most from Africa. Each of these workshops had a different group of approximately 50-100 learners each day.

It is important to note that Wood and Cecire had the help of ASP Lecturers in the workshops, particularly in the second week. These Lecturers, all accomplished physicists, added information, influenced the offerings, answered questions, and were generally helpful. They also added authenticity to the proceedings because they could discuss contemporary physics research and practice first-hand with the participants.

Another feature of the second week were two hands-on sessions Wood and Cecire offered for the ASP students. The first, on December 5, was a Cosmic Ray workshop in which students assembled a QuarkNet Cosmic Ray Detector and took data and then experimented with smaller Cosmic Watch detectors under evaluation by QuarkNet. The second, on December 8, was an ATLAS masterclass in which students analyzed events from the ATLAS detector at the Large Hadron Collider at CERN and examined the results.

This report will detail the above-described activities and offer informal evaluative comments and conclusions.

**Teachers Program**
*Schedule.* The schedule in Figure 1 below indicates the general plan of the Teachers Program.

# Report on Notre Dame Contribution to the African School of Physics 2022

## High School Teachers Program Details

| | Nov 28 | Nov 29 | Nov 30 | Dec 01 | Dec 02 |
|---|---|---|---|---|---|
| 08:30-09:00 | Guests of Honor addresses | News and Information | | | |
| 09:00-10:30 | 100 Years of Physics & Future / plenary with students | Physics for Sustainable Development for students<br><br>Newton's Laws for Teachers | Registration, curriculum discussion, and introductory activities / Electromagnetism / parallel with students | Neutrinos with MINERvA / parallel with students | Small detector Labs for teachers<br>ATLAS or CMS masterclass measurement |
| 10:30-11:00 | | BREAK | | | |
| 11:00-12:30 | Standard Model of Particle physics for students<br><br>Energy and Power for Teachers | Standard Model of Cosmology for students<br><br>Newton's 2nd Law for teachers | Review of and activities related to particle and astroparticle physics / Electrostatics / parallel with students | Higgs@10, part I/ parallel with students | Small Detector Labs for teachers<br>Curriculum Connections and School Implementation |
| 12:30-14:00 | | LUNCH | | | |
| 14:00-15:30 | Radiation Physics & particle interactions with matter for students<br>Wave, sound and light for teachers | Basics of Accelerators for students<br><br>Photoelectric effect for teachers | Muons with MINERvA Electricity / parallel with students | Higgs@10, part II/ parallel with students | Small Detector Labs for teachers |
| 15:30-16:00 | | BREAK | | | |
| 16:00-17:30 | TEACHERS MAY LEAVE OR JOIN STUDENTS PROGRAM | | | | |
| 17:00-18:30 | | BREAK | | | |
| Evening Colloquia / Public lectures | TEACHERS AND THE GENERAL PUBLIC MAY ATTEND EVENING PUBLIC LECTURES | | | | |

*Figure 1. High School Teachers program schedule of activities.*

The sessions at 09:00 on November 28 and 29 were either plenaries attended by both ASP students and the teachers or teacher-only sessions related to physics topics in the South African Physical Science standards. On November 30, local Physical Science sessions were scheduled to be offered in parallel with contemporary physics (titles in red); however, teachers and their supervisors decided to add later afternoon sessions to cover these so that all teachers could attend the contemporary physics sessions facilitated by Wood and Cecire. Contemporary physics offerings continued on December 1 and December 2. Due to a miscommunication, most teachers had understood that they could travel home on December 2 and were offered two alternative electronics sessions that morning in parallel with the offerings by Cecire and Wood. Some stayed for the first electronics session but almost all had started for home by the second session. The December 2 contemporary physics sessions were in a different and more distant building than any that the teachers had visited all week and, consequently, none made it to that location. Cecire and Wood used the time for planning.

*Participation.* A group of 76 South African teachers, most from the Eastern Cape province of South Africa, participated in the Teachers Program. The majority were teachers of Physical Science, the coursework in the South African curriculum that covers both physics and chemistry at the high school level. Wood and Cecire attended several sessions with the teachers on November 28 and 29 but also took time to prepare for the sessions they facilitated that began on November 30. At that point, Wood and Cecire gave the teachers an online supplementary registration form, which 51 filled out. Of these, 44 reported that they were high school teachers, one was an elementary school teacher, and the remaining six were administrators of one kind or another.

# Report on Notre Dame Contribution to the African School of Physics 2022

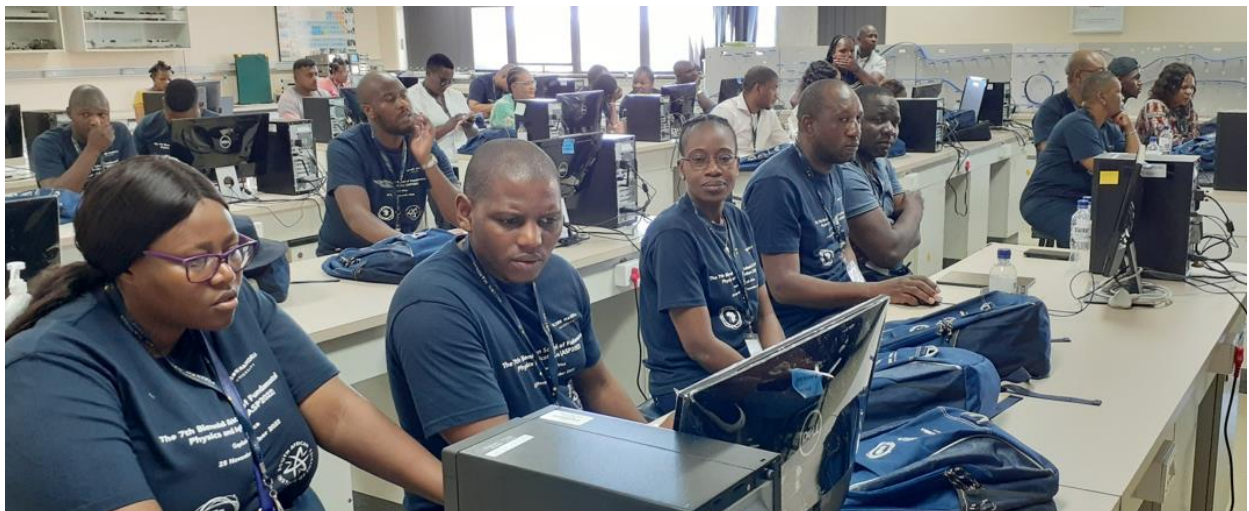

*Figure 2. Teacher Program participants.*

*Days 1 and 2.* The local Physical Science sessions were intended to be hands-on but the requisite equipment was unavailable so they were converted to discussions of how to present and how learners may respond to these presentations. The discussions were quite spirited, especially as the teachers from many different schools became more comfortable with each other. The teachers were also quite insightful as to how learners might respond. In one session, the facilitators explained that the Physical Science standards covered only the vertical case of projectile motion; the teachers were later shown a straight-line velocity-time graph of this motion. When asked about the learner reaction, one teacher took on the learner role and said, "You told me the motion was only vertical but now you are showing me a slope!" The facilitators from the University of Limpopo worked hard to make the mostly-discussion sessions work and were unfailingly cheerful and did a very good job of connecting with the teachers.

*Days 3, 4, and 5.* On November 30, Cecire and Wood facilitated activities for the teachers from the QuarkNet Data Activities Portfolio [5]. These are activities for high school learners that help them to learn about particle physics or use particle physics to understand topics in the high school physics curriculum. The teachers took on the role of learner to perform and understand each activity and then were able to discuss the activities as teachers. Wood and Cecire strived, where practical, to connect the activities to South African Physical Science standards. They were, overall, quite pleased to get to hands-on work that day.

The activities for November 30, all found in the QuarkNet Data Activities Portfolio, included:
- Shuffling the Particle Deck, in which small groups were given cards representing fundamental particles with various characteristics of the particles printed on the cards. Groups organized the cards based on those characteristics. See Figure 3.
- Human Particle Tricks (not in the Data Portfolio but often used in QuarkNet), in which participants take the roles of elementary particles and act out particle interactions, such as two protons colliding to create a Z boson, which in turn decays into a muon and anti-muon pair.
- Rolling with Rutherford, where small groups take turns rolling one marble ten times at four other fixed, but otherwise identical, marbles, count the number of hits, make a histogram to find the most probable number of hits out of ten, and then use a simple statistical argument to calculate the diameter of a marble. See Figure 4.
- Mean Lifetime Part I: Dice, where participants roll dice multiple times until they land on a specific number, at which point the die is deemed "decayed". The number of dice left after each roll is recorded and put on a plot of dice remaining as a function of the number of rolls. They get an exponential decay and determine the lifetime and half-life of the dice.
- Mean Lifetime Part II: MINERvA uses muon tracks in the MINERvA neutrino detector at Fermilab. The teachers use the actual muon data from a Fermilab experiment to determine how long each muon lasts in the detector before it decays and how much energy the electron decay product has in each decay. Similar to the previous activity, they determine the muon lifetime with good accuracy. They also use the electron energy to "work backwards" and determine the muon mass, again with good accuracy.

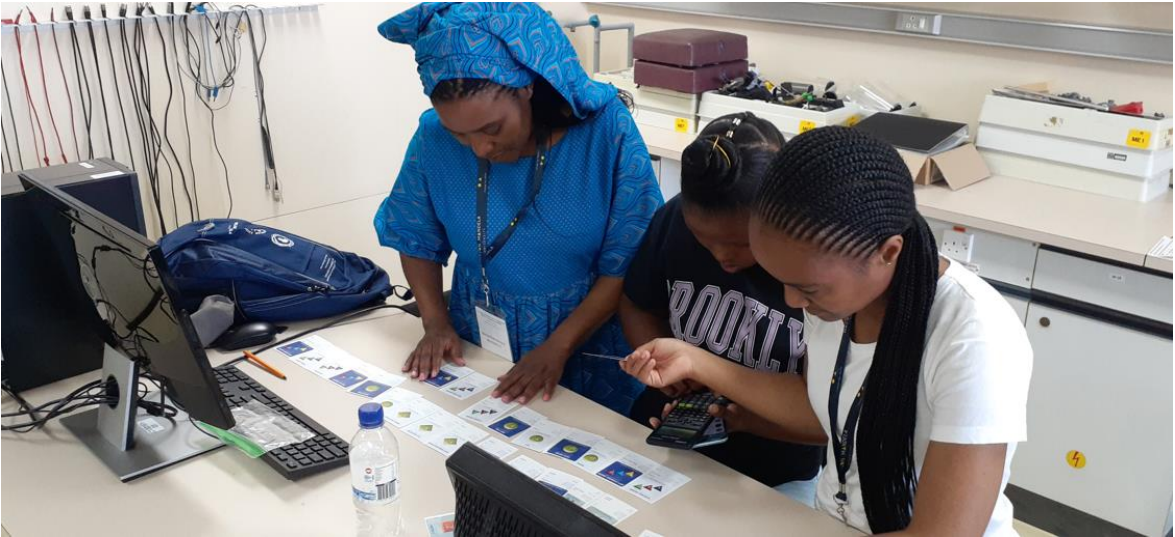

*Figure 3. Teachers organize cards in Shuffling the Particle Deck.*

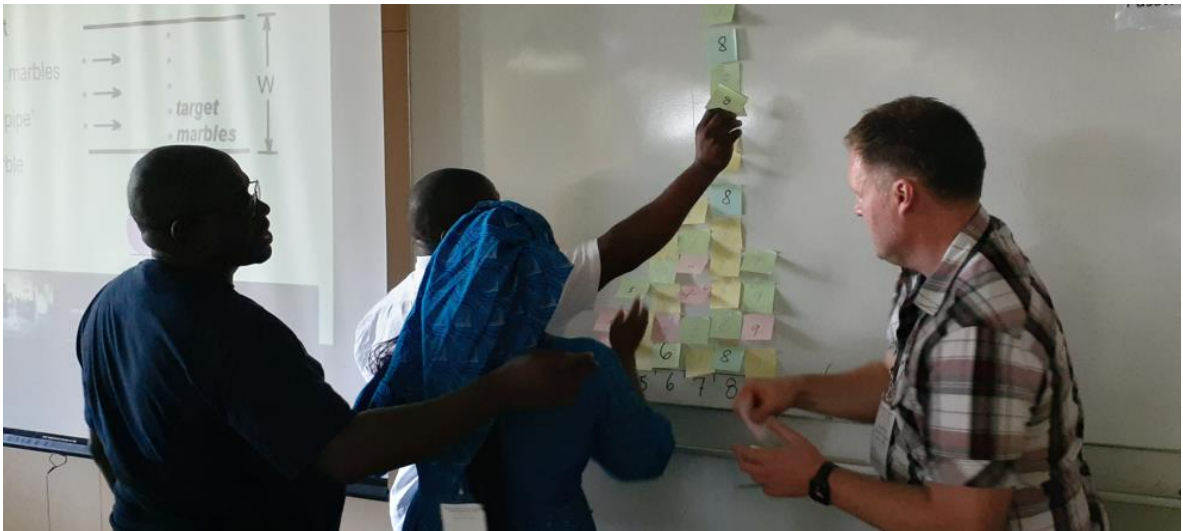

*Figure 4. Creating the histogram in Rolling with Rutherford.*

**Report on Notre Dame Contribution to the African School of Physics 2022**

On December 1, Cecire and Wood resumed with the teacher group. The work of the day consisted of two more intensive activities:

- MINERvA Neutrino Masterclass [6], which is one of the original contributions of QuarkNet to International Masterclasses. Participants use an event display to evaluate the momenta of particles that emerge from the carbon nucleus after it interacts with a neutrino from a beam produced in the Fermilab accelerator complex. From this, they discover the motion of neutrons inside the nucleus. Due to time constraints, this masterclass was introduced but not completed.
- Higgs@10 [7], a workshop created by QuarkNet to help teachers learn about the Higgs boson on the 10th anniversary of its discovery. Teachers navigated this successfully, though the calculations at the end proved to be somewhat intense.

This day was successful overall but not as much as the previous day, since the work was more intensive and time-consuming. If we had focused on only one of the two main activities it would have been better.

December 2 was the day that teachers believed they were to go home. A new set of activities for the teachers took up the rooms where they had met with Cecire and Wood previously; the QuarkNet-inspired activities were moved to a new, unfamiliar location farther from the main meeting room. As a result, it became a planning day for Cecire and Wood to refine the Learners Program for the next week.

Teacher Evaluation. Cecire sent a survey to the teachers on January 30, 2023. [9] Only eight teachers responded, likely due to the lateness of the survey and the press of work at their schools. Nevertheless, the partial results that did come in will help to guide future efforts. In rating favorite parts of the November 30 and December 1 workshops, there was wide acceptance of most activities, with only some parts of Higgs@10 lagging, as seen in Figure 5.

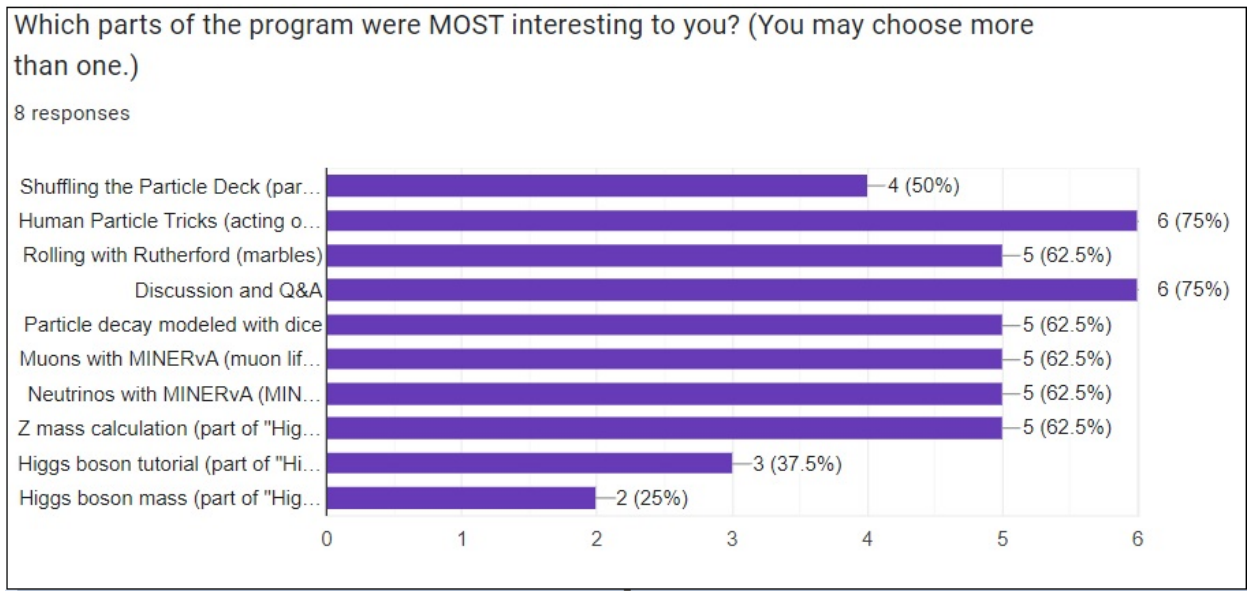

*Figure 5. Favorite activities of teachers in the workshop.*

Tracking this, the least favored activity according to 50% of the respondents was Higgs boson mass calculation. This was probably the most intensive part, requiring teachers who may have been unfamiliar with spreadsheet calculations to do mathematical calculations of a large dataset in a spreadsheet.

The overall rating of the two days was good, according to the respondents.

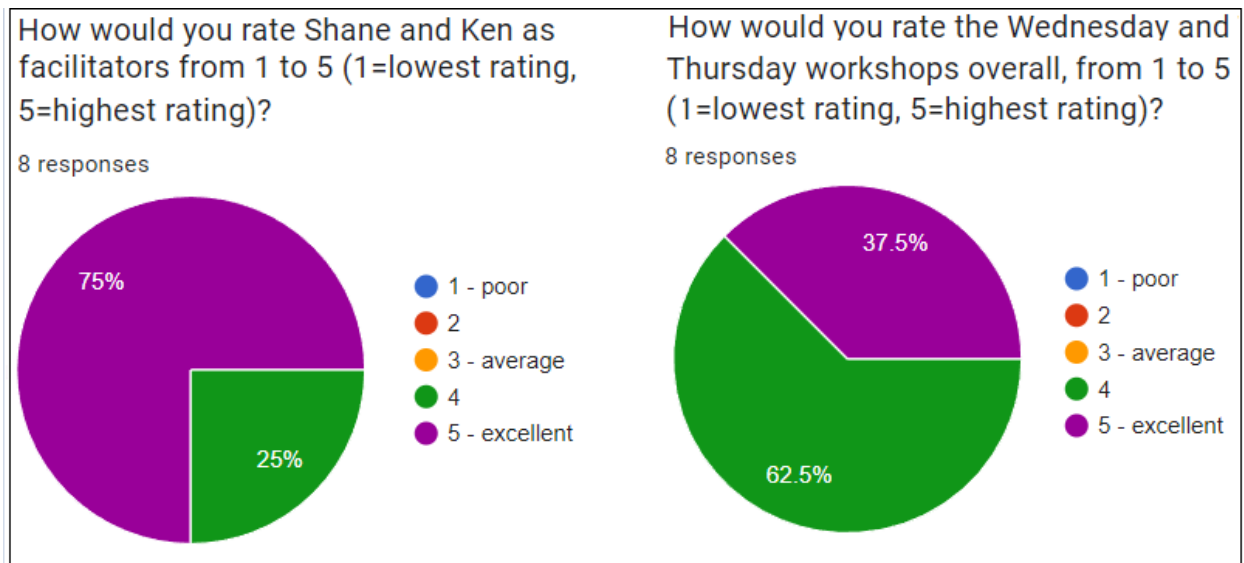

*Figure 6. Rating of facilitators (left) and of workshop (right).*

Selected teacher responses to queries:

- *What, if anything, might you bring from Wednesday and Thursday of the Teacher Program into your own classroom?*
    - Everything.
    - Showing learners smart ways of calculations and rolling dice.
    - The teaching strategies employed by the facilitators was excellent.
    - Practical demonstration.
    - Shuffling the particle deck.
    - Human particle tricks.
    - Learners need to be introduced to particle physics in the simplest way possible.
    - Teacher, learner involvement.
    - Role modeling & participation of teachers and the way concepts were integrated through fun.
- *Do you have any other comments?*
    - I would love to get videos of how teachers tackle certain topics in physics there especially those that are so abstract to learners. I would also love to experience a real classroom atmosphere on that side of the Earth.
    - I also learned a trick on how to sort a large class
    - Will forward email
    - It was interesting

- Energy displayed by the presenters showed passion and stimulated interest even though we were hearing some terms for the first time hahaha like Muons…

**High School Learners Program**
The High School Learners Program took place at two venues outside of the Nelson Mandela University main campus. The first two days, December 5 and 6, were at the Nelson Mandela University Missionvale Campus, about 25 kilometers from the Main Campus in Summerstrand. The remaining three days were at the Nelson Mandela Bay Science and Technology Centre in Kariega (Uitenhage), an additional 20 km from Summerstrand. The program consisted of a morning workshop each of the five days with a different group of learners with their teachers from nearby schools. The format was nearly identical each day with a varying number of participants that, in the end, added up to 231 learners. For the most part, then, we will treat the program as a whole.

The daily schedule varied some due to transport issues for the facilitators and different arrival times of learners from different schools but basically followed this agenda:

09:00 Arrival and introduction
09:30 Shuffling the Particle Deck (Figure 7)
10:00 Begin Rolling with Rutherford (Figure 8)
10:30 Break with exhibits (Figure 9)
11:00 Finish Rolling with Rutherford (Figure 10)
11:30 Human Particle Tricks (Figure 11)
12:00 Q&A with physicists (Figure 12)
12:30 Dismissal

Slides and videos were used at various times throughout the workshop to illustrate key points. One video was used to inform learners about basic research in South Africa through the Square Kilometer Array radio observatory. A QuarkNet cosmic ray detector was available during the break for inspection and demonstration.

A key part of the workshop was the presence of ASP physicist lecturers, mostly from Africa, who assisted in facilitation and were the objects of the Q&A, which was introduced as, "Ask the physicists anything."  Working on activities with assistance from physicists and the ability to ask questions of any sort to those physicists was a unique opportunity for learners who may never have met a scientist before.

Learners were enthusiastic and admirably well-behaved, even as schedule glitches made proceedings difficult at times [9]. It was clear that some groups were better prepared in terms of physics background than others and it may have also been true that some groups had more or less proficiency in English, which was their second language behind, mostly, Xhosa. (Students whose first language was Afrikaans would have had an easier time of this as that language and English have many similarities.)

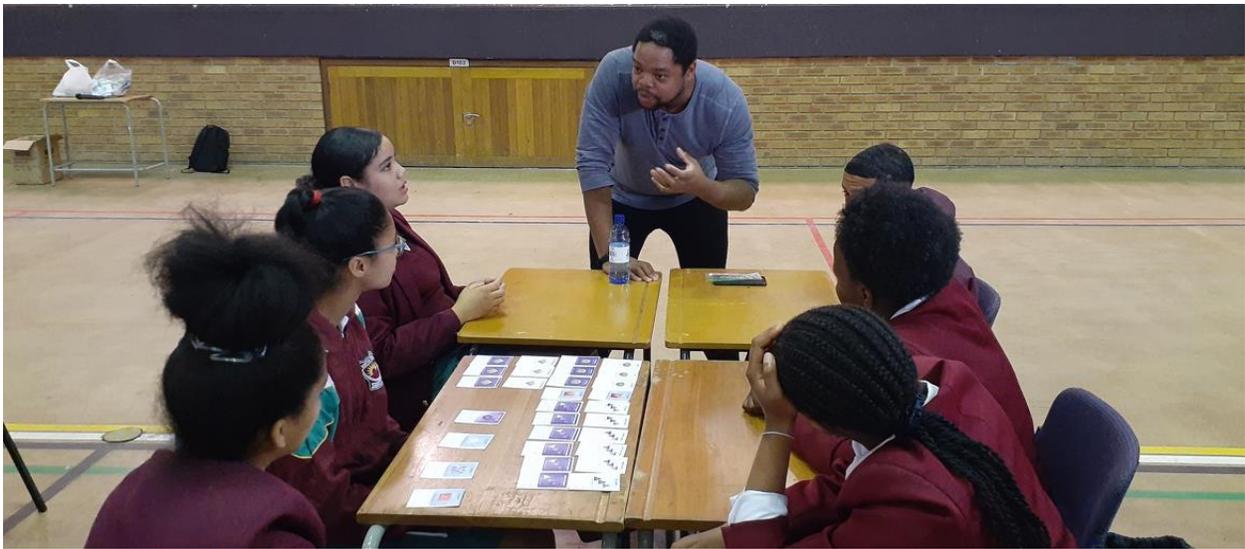

*Figure 7. Gopoleng Mohlabeng (University of California Irvine, originally from South Africa) discusses the particle cards and the Standard Model with learners in Missionvale.*

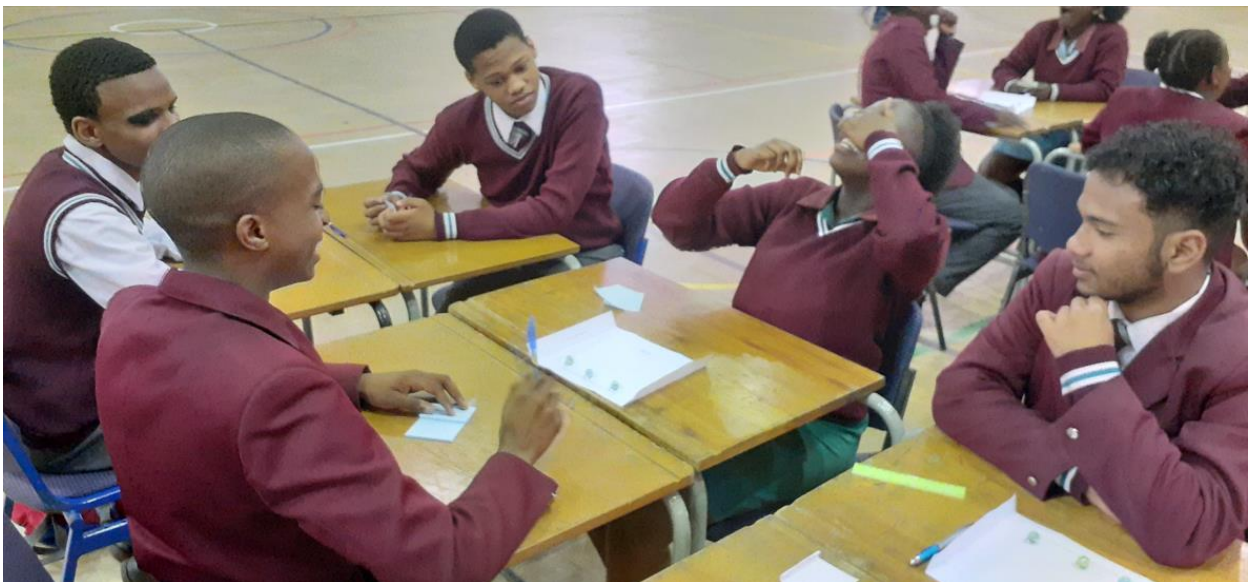

*Figure 8. A learner in Missionvale works extra hard to make her marble roll random to get the statistics right.*

**Report on Notre Dame Contribution to the African School of Physics 2022**

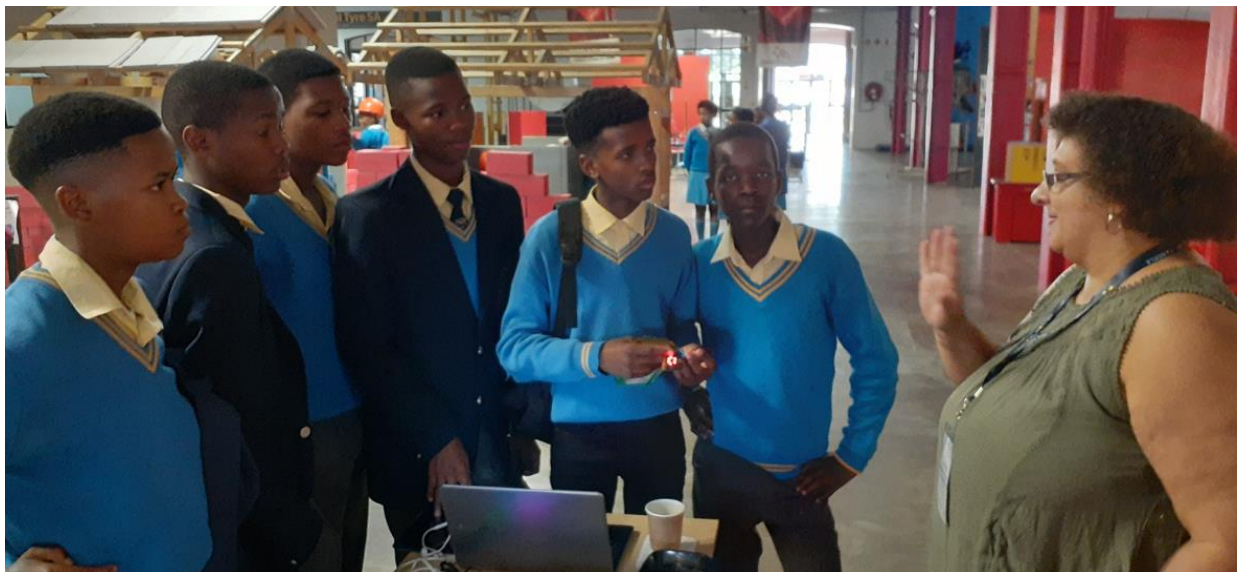

*Figure 9. Mary Bishai of Brookhaven National Laboratory (originally from Egypt) talks physics during a break with learners in Kariega.*

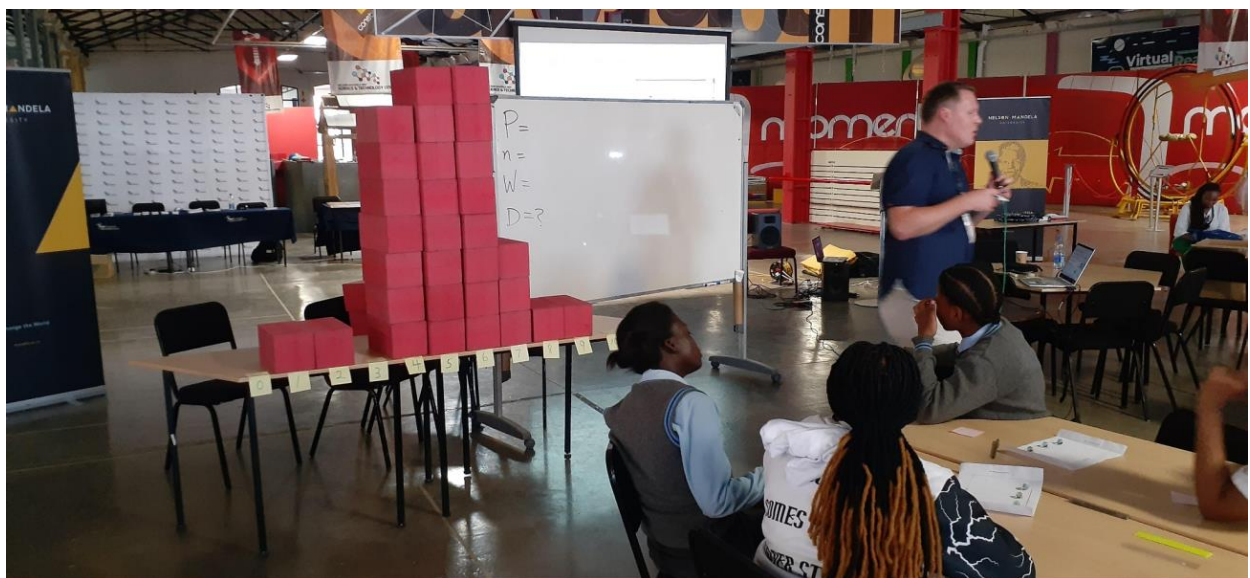

*Figure 10. Shane Wood and learners discuss the Rolling with Rutherford histogram made with foam blocks in the Nelson Mandela Bay Science and Technology Centre in Kariega.*

**Report on Notre Dame Contribution to the African School of Physics 2022**

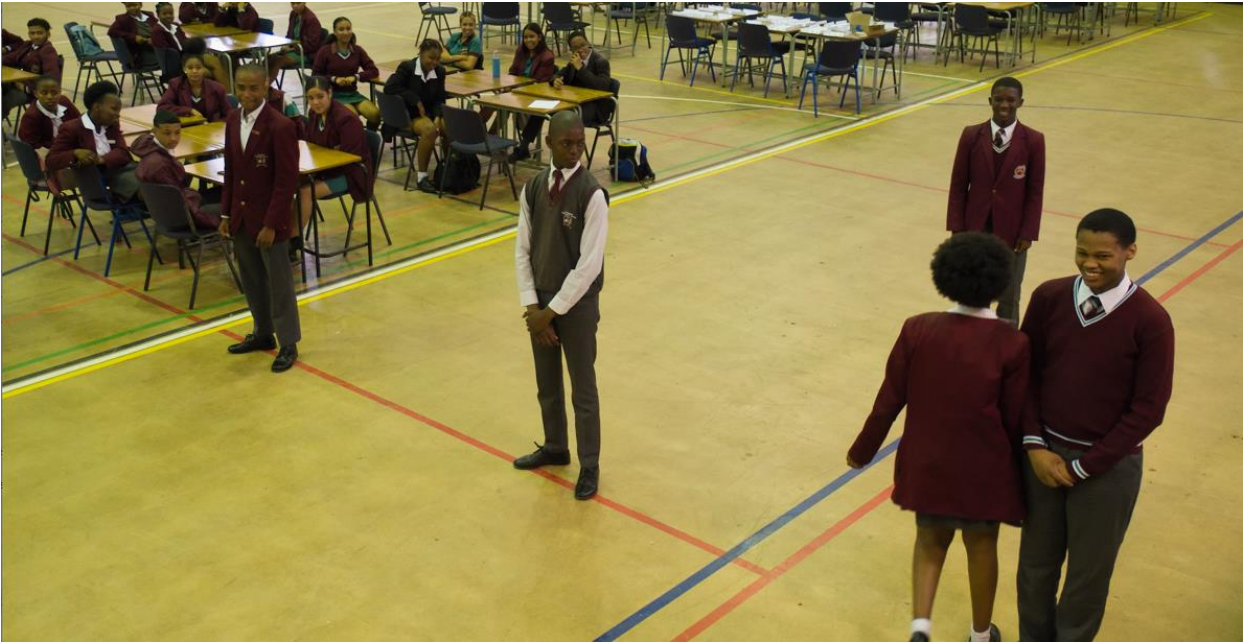

*Figure 11. Human particle interaction performed by learners in Missionvale.*

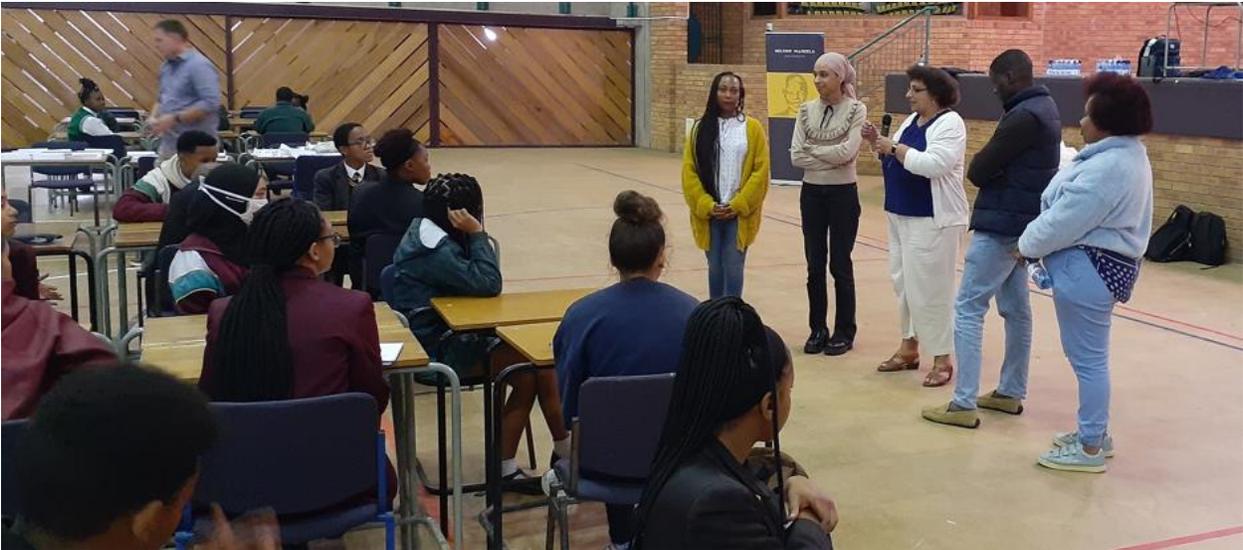

*Figure 12. Physicist Q&A in Missionvale with (L to R) Ann Njeri, University of Manchester (originally from Kenya), Mounia Laassiri, University of Helsinki (originally from Morocco), Mary Bishai, Diallo Boye, Brookhaven National Laboratory (originally from Senegal), and Hashe Nobom, Nelson Mandela University (from South Africa).*

**Report on Notre Dame Contribution to the African School of Physics 2022**

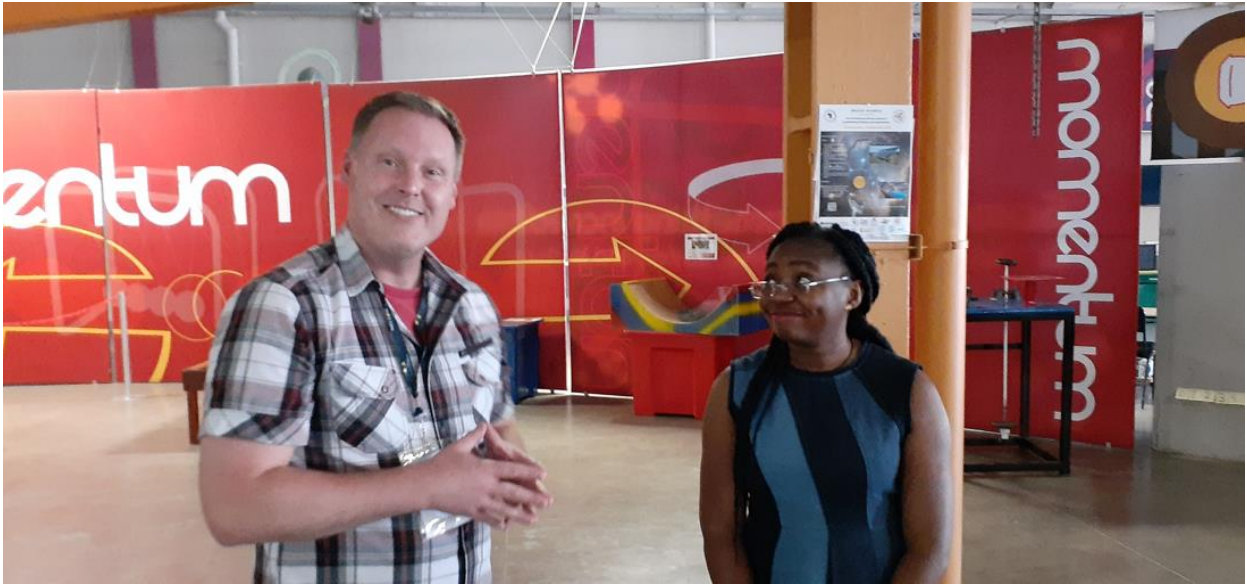

*Figure 13.Shane Wood and Esmeralda Yitamben of Merck (originally from Cameroon), share satisfaction with the workshop in Kariega.*

**University Student Program**

Wood and Cecire gave two classes for ASP students in parallel sessions. One class was on December 5 and the other was on December 8, both for 90 minutes in the afternoon.

The first class was on cosmic ray detectors. Students put a QuarkNet cosmic ray detector [10] together and set it up to take data on a computer. This included setting operating voltages, adjusting the location of the GPS antenna to where it could get a fix on satellites, and recording the data on the computer. See Figure 14 below. Once they were done with this, the students were given Cosmic Watches, small detectors that had been built at Notre Dame, to try out experiments based on the design from MIT [11]. These took a bit more time than expected to set up and some did not work properly - they are still in the testing phase - but students were able to see some of the possibilities. Overall, the students had a good hands-on experience with actual particle detectors and learned about the main components of such detectors. About ten students attended.

**Report on Notre Dame Contribution to the African School of Physics 2022**

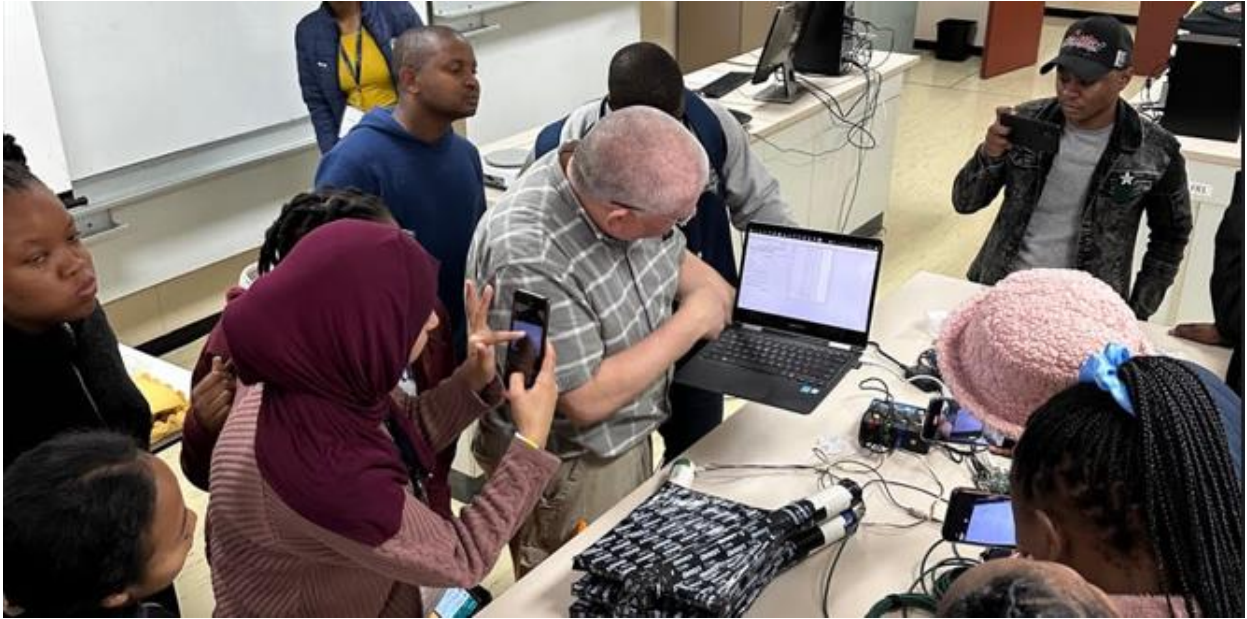

*Figure 14. Cecire and students begin taking data from the QuarkNet detector.*

The second class was a streamlined ATLAS Z-path masterclass [12] with four students. In this activity, each student was assigned a dataset of events from the ATLAS particle detector in the Large Hadron Collider at CERN. They examined these visually, determined the sort of event each was, and recorded energy and momentum data from ATLAS that allowed them to measure the mass of the parent particle in each event; these results were aggregated to produce a histogram that revealed parent particles at specific mass peaks, shown in Figure 15 below. Of the four who came to the masterclass, three became newly interested in pursuing particle physics in graduate school after analyzing the ATLAS data.

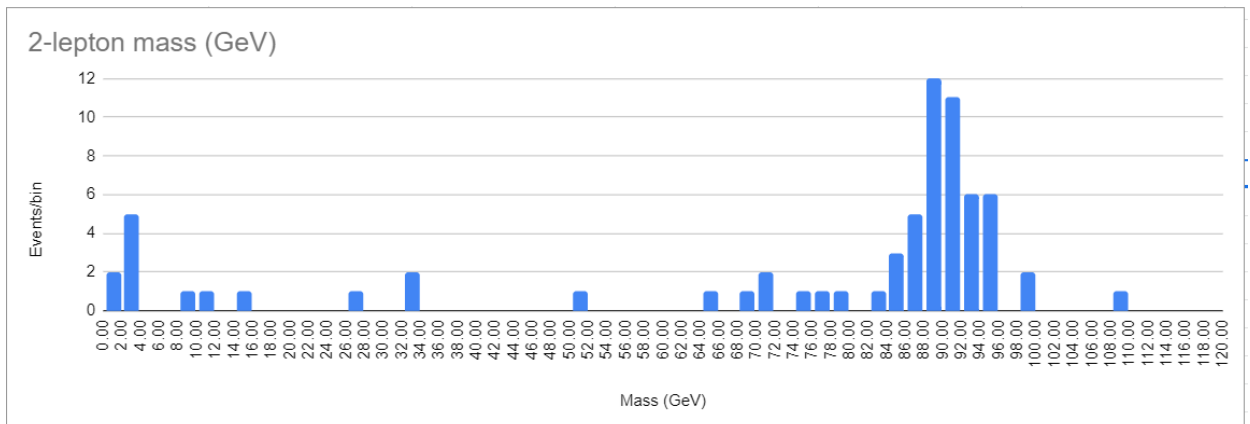

*Figure 15. ATLAS results from the ASP2022 masterclass. A small peak for the J/Psi meson is just visible near 3 GeV mass while a prominent peak for the Z boson is near 90 GeV.*

**Conclusions**

The Notre Dame QuarkNet participation was successful in modeling good physics pedagogy and the methods of particle physics to teachers and learners in South Africa who have not had access to these in the past. Teachers seemed to really enjoy the more hands-on and data-oriented approach to physics education and several expressed in various ways how it might influence their teaching. For learners, this was often their first exposure to the quantum world of fundamental particles as well as their first chance to interact directly with physicists and to see them as real people. These interactions, we believe, open a new world to those learners.

The ASP students benefited from the topics and the approach Cecire and Wood brought to them. ASP is rather lecture-oriented as a whole: the cosmic ray and masterclass activities, along with microcontroller experiments brought by retired CERN physicist Ulrich Raich, enabled the students to experience experimental physics a little more directly.

The Lecturers who assisted us also added constructive comments during and after the workshops. The most prominent was that we should try to add more content from other fields than particle physics, for example astrophysics or material science, to the Learner Program. We agree with the constraints that this material must fit into the overall workshop without straining the schedule and that we use well-vetted activities of quality similar to the QuarkNet activities we already use. We might be able to create such activities over the next year or so. To fit in, we might vary the offerings on each day of the Learner Program, meaning that some student groups will have a different experience than other groups. We have collected written communications that we solicited after ASP 2023. [13]

Another sort of success has been built over time through involvement in the ASP program since 2016 and continues into this most recent iteration: it is the value of collaboration. Wood and Cecire were enriched by working with the organizers from Nelson Mandela University and the ASP lecturers, especially those who helped with the Teacher and Learner Programs. Not only have the names of QuarkNet and Notre Dame come in front of these colleagues but there are also real results. For example, two of the lecturers, Chilufya Mwewa and Benard Mulilo, who are ASP alumni, are working with Cecire and International Masterclasses to build the first-ever ATLAS masterclass in Zambia, scheduled for March 27, 2023. Notre Dame QuarkNet is known in ASP for its leadership of Teacher and Learner Programs and the activities it brings; in this capacity, Cecire is a member of the International Organizing Committees for ASP and its sister program, the African Conference on Fundamental Physics and Applications (ACP). [14]

We recommend that the relationship between Notre Dame, QuarkNet, and the African School of Fundamental Physics and Applications be continued and even deepened by the participation, if practical, of a member of the Notre Dame Department of Physics and Astronomy as a regular lecturer.

**Supplementary Material**
All of the images in this report are kept in a file on the QuarkNet server. Most of these are "full size". The same directory contains a report by a lecturer, a report by a Nelson Mandela University staff member, and collected comments by lecturers. [13] The official ASP 2022 Activity Report is available on arXiv [15].

**Appreciation**
Kenneth Cecire and Shane Wood would like to thank Professor Mitchell Wayne, Associate Dean Michael Hildreth, Dean Santiago Schnell, QuarkNet, and the Notre Dame College of Science for going "above and beyond" to make this program possible. We also extend great thanks to the team at Nelson Mandela University - Azwinndini Muronga, Jade Alexander, Sisipho Ngesi, Dolly Ntintili, Reatile Mosia, Andre Venter, Brian Masara, Nobom Hashe, and

# Report on Notre Dame Contribution to the African School of Physics 2022

Albert Hsien-Chang Liu - all unfailingly helpful, kind, and cheerful, to the Lecturers who joined the outreach programs and made a tremendous difference, and to Ketevi Assamagan for his leadership, guidance, and thoughtfulness.

**References and Notes**

1. https://www.africanschoolofphysics.org/
2. https://quarknet.org
3. https://www.africanschoolofphysics.org/asp-2016/
4. https://web.quarknet.org/files/asp/SouthAfrica_PhysicalScience_2011.pdf
5. https://quarknet.org/data-portfolio
6. https://indico.fnal.gov/event/22340/
7. Example agenda from QuarkNet workshop at Syracuse University (middle column of table): https://quarknet.org/content/2022-syracuse-university-workshop
8. https://forms.gle/VnGA4d22E5Q4Xr1G8
9. On the final day in Kariega, the temperature reached 40 degrees Celsius (about 100 degrees Fahrenheit) with only fans for cooling. The learners never complained and, indeed, most continued to wear their school uniform sweaters. We ended the workshop a bit early that day.
10. Users Manual: https://quarknet.org/sites/default/files/cf_6000crmdusermanual-small.pdf
11. http://www.cosmicwatch.lns.mit.edu/
12. https://atlas.physicsmasterclasses.org/en/zpath.htm
13. https://web.quarknet.org/files/asp/gqeberha2022_report/
14. https://indico.cern.ch/event/1229551/
15. Assamagan, Ketevi, Acharya, Bobby, Cecire, Kenneth, Darve, Christine, Ferroni, Fernando, Gray, Julia Ann, and Muronga, Azwinndini, *Activity Report on the Seventh African School of Fundamental Physics and Applications (ASP2022)*, arXiv:2302.13940 [physics.ed-ph], https://arxiv.org/abs/2302.13940.